\documentclass[appendixfloats]{emulateapj}
\usepackage{graphicx}
\usepackage{graphics}
\usepackage{color}
\usepackage{epsfig}
\usepackage{url}
\usepackage{hyperref}
\usepackage{breakurl}
\usepackage{scalerel}

\shortauthors{J. T. Li et al.}
\shorttitle{CGM-MASS I: XMM-Newton Large Project and NGC~5908}

\begin{document}

\title{The Circum-Galactic Medium of MASsive Spirals I: Overview and a Case Study of NGC~5908}

\author{Jiang-Tao Li\altaffilmark{1}, Joel N. Bregman\altaffilmark{1}, Q. Daniel Wang\altaffilmark{2}, Robert A. Crain\altaffilmark{3}, and Michael E. Anderson\altaffilmark{4}} 

\altaffiltext{1}{Department of Astronomy, University of Michigan, 311 West Hall, 1085 S. University Ave, Ann Arbor, MI, 48109-1107, U.S.A.}

\altaffiltext{2}{Department of Astronomy, University of Massachusetts, 710 North Pleasant St., Amherst, MA, 01003, U.S.A.}

\altaffiltext{3}{Astrophysics Research Institute, Liverpool John Moores University, IC2, Liverpool Science Park, 146 Brownlow Hill, Liverpool, L3 5RF, United Kingdom}

\altaffiltext{4}{Max-Planck Institute for Astrophysics, Karl-Schwarzschild-Stra$\rm\beta$e 1, 85748 Garching bei M\"{u}nchen, Germany}


\nonumber

\begin{abstract}
The {\color{red}C}ircum-{\color{red}G}alactic {\color{red}M}edium of {\color{red}MAS}sive {\color{red}S}pirals (CGM-MASS) is a project studying the overall content, physical and chemical properties, and spatial distributions of the multi-phase circum-galactic medium (CGM) around a small sample of the most massive ($M_*\gtrsim2\times10^{11}\rm~M_\odot$, $v_{\rm rot}>300\rm~km~s^{-1}$) isolated spiral galaxies in the local Universe. We introduce the sample and present a detailed case study of the \emph{XMM-Newton} observation of the hot gas halo of NGC~5908. After data calibration, point source removal, and background analysis, we find that the diffuse soft X-ray emission of NGC~5908 is significantly more extended than the stellar light in the vertical direction. The 0.5-1.25~keV radial intensity profile tracing hot gas emission can be detected above the background out to about $2^\prime$, or $30\rm~kpc$ from the nucleus. The unresolved soft X-ray emission can be characterized with a $\beta$-model with a slope of $\beta\approx0.68$. The unresolved 0.5-2~keV luminosity within $1^\prime$ is $6.8\times10^{39}\rm~ergs~s^{-1}$, but young stellar sources may contribute significantly to it. Assuming a metallicity of 0.2~solar, an upper limit (without subtracting the very uncertain young stellar contribution) to the mass of hot gas within this radius is $2.3\times10^9\rm~M_\odot$. The cooling radius is $r_{\rm cool}\approx25\rm~kpc$ or $\approx0.06r_{\rm 200}$, within which the hot gas could cool radiatively in less than 10~Gyr, and the cooling of hot gas could significantly contribute in replenishing the gas consumed in star formation. The hot gas accounts for $\approx1.9\%$ of the baryon detected within the cooling radius. By comparing NGC~5908 to other galaxies, we find that its X-ray luminosity per stellar mass is consistent with lower-mass non-starburst field spiral galaxies. However, a large scatter in hot gas soft X-ray emissivity is indicated for spiral galaxies with $M_*\gtrsim2\times10^{11}\rm~M_\odot$.
\end{abstract}


\section{Introduction}\label{sec:Introduction}


In the $\rm \Lambda CDM$ galaxy formation theory (e.g., \citealt{White91}), the circumgalactic medium (CGM) around present-day spiral galaxies serves as a reservoir from which the galaxies acquire baryons to continue star formation and build up galactic disks. The CGM also stores the kinetic energy and chemically enriched materials deposited by galactic feedback. A potentially powerful probe of the CGM is the diffuse soft X-ray emission from hot gas (e.g., \citealt{Crain10}). Indeed, it has been observed in and around many galaxies (e.g., \citealt{Li13a}). However, the X-ray emission of the CGM could arise not only from the feedback of active galactic nuclei (AGN), stellar winds, and supernovae (SNe) (e.g., \citealt{Strickland04,Li13b,Wang16}), but also from the accretion shock heating and gravitational compression of the intergalactic medium (IGM) (e.g., \citealt{White91,Benson00}). The relative importance of these two potentially inter-related mechanisms likely depends on a galaxy's mass, as well as other properties such as the star formation rate (SFR) and the environment. At present, these dependences still need to be carefully examined.


Statistical studies of a sample of galaxies should, in principle, allow us to disentangle the complex relations of CGM properties with various other galaxy parameters (e.g., \citealt{Strickland04,Tullmann06,Li11,Li13b,Wang16}). However, the interpretation of the well established relationships is not straightforward. For example, the observed linear $L_{\rm X}-{\rm SFR}$ relation is thought to be a result of the scaling of both $L_{\rm X}$ and SFR on the galaxy mass \citep{Crain10,Wang16}, and it is even suggested that the star formation could play a negative role in producing the X-ray emission for massive galaxies with existing hot coronae \citep{Li13b,Wang16}. Furthermore, even the nature of the X-ray emission itself is uncertain (e.g., \citealt{Strickland02}). It has been suggested that the charge exchange at boundaries of hot and cool gases can sometimes be significant in addition to the optically-thin thermal plasma emission in soft X-ray bands (e.g., \citealt{Liu12,ZhangS14}). It is also possible that a significant fraction of the unresolved soft X-ray emission is comprised of various stellar sources, such as young stars (e.g., \citealt{Sazonov06}), Cataclysmic Variables (CVs) and coronal active binaries (ABs) (e.g., \citealt{Revnivtsev08}), as well as low- and high- mass X-ray binaries (LMXBs and HMXBs; e.g., \citealt{Gilfanov04,Mineo12}).

In spite of the above difficulties in both theoretical and observational efforts, there are some pilot works directly comparing the X-ray observations to theories. These early comparisons are subject to poorly understood systematics in terms of both models and observations, because of the limited sample size and the non-uniformity of the observational and/or numerically simulated data (e.g., \citealt{Benson00,Toft02,Rasmussen09}). These works often predict a significant X-ray emitting corona around present day spiral galaxies from numerical simulations, but the expected soft X-ray luminosity often exceeds what is observed in nearby galactic halos. 

\citet{Li14} compared the X-ray observations from \citealt{Li13a,Li13b}'s sample to a comparable sample of simulated galaxies drawn from the Galaxies-Intergalactic Medium Interaction Calculation (GIMIC; \citealt{Crain09,Crain10}). After carefully decomposing different X-ray emission components, they rescaled the X-ray luminosity to the same radial range of $r=(0.01-0.1)r_{200}$, where $r_{200}$ is the virial radius enclosing a sphere with a mean density of 200 times the critical density of the Universe. They found that the GIMIC simulations could roughly reproduce the observed X-ray luminosity range near the optical extent of non-starburst $L^\star$ galaxies. However, some significant discrepancies between observations and simulations still exist, which could come from the large uncertainties in both observations and simulations. 

We herein present a brief introduction of our new \emph{XMM-Newton} survey of nearby massive spiral galaxies, and an initial case study of NGC~5908. The goal of this study is to lay out a detailed data calibration and analysis procedure that is intended to be used for the entire sample. The remaining part of this paper is organized as follows: In \S\ref{section:Sample}, we introduce our sample selection criteria of the most massive spiral galaxies in the local Universe and the \emph{XMM-Newton} large project approved in AO-13. We present detailed data reduction in Appendix \S\ref{subsection:DataReduction} and analysis of the X-ray bright point-like sources and prominent extended soft X-ray features in Appendix \S\ref{section:PointSrcExtendFeature}. Scientific analysis of NGC~5908 is presented in \S\ref{section:CaseStudyN5908}. The results are discussed in \S\ref{section:ResultsDiscussion}. We finally summarize the major conclusions in \S\ref{section:Summary}. Data calibration of other galaxies in the \emph{XMM-Newton} large project, as well as further analyses and scientific discussions, will be presented in follow-up companion papers.


\section{The CGM-MASS project}\label{section:Sample}


The {\color{red}C}ircum-{\color{red}G}alactic {\color{red}M}edium of {\color{red}MAS}sive {\color{red}S}pirals (CGM-MASS) is a project studying the overall content, physical and chemical properties, and spatial distributions of the multi-phase circum-galactic medium (CGM) around a sample of the most massive spiral galaxies in the local Universe. We choose to focus on spiral galaxies, because elliptical galaxies are most commonly found in dense, clustered environments where it is difficult to discriminate between the establishment of a hot CGM and a hot IGM. As the first step, we are now collecting and analyzing the X-ray data to study the hot CGM. We aim at collecting enough photons at large galactocentric radii, in order to measure or tightly constrain not only the coronal X-ray luminosity, but also the hot gas temperature and metallicity, and hence some derived parameters such as the gas density, thermal energy, and total hot baryon content of the galaxies. \emph{XMM-Newton}, with its high sensitivity and large field of view (FOV), is thus optimized for this purpose.

\subsection{The lack of high-quality X-ray observations of massive spiral galaxies}\label{subsec:LackXrayData}

Typically, the X-ray emitting corona of an isolated spiral galaxy is comprised of two components. The X-ray emission in the \emph{inner} halo close to the immediate vicinity of a galaxy's optical extent is most likely related to the stellar feedback (e.g., \citealt{Li13b,Li15a,Wang16}). In contrast, there is also expected to be a more spherically symmetric and more extended hot gas component in the \emph{outer} halo which is typically beyond the stellar disk and bulge and the close vicinity of the galaxy dominated by the galactic outflows. This component, as shown in numerical simulations, is more dilute and is more likely related to the gravitational processes dominating the hierarchical structure formation. These two components, however, are often \emph{not} explicitly separated and could mix with each other. In general, the outer halo has low metallicity and density, so faint in soft X-ray, but could dominate the total hot gas mass in a galactic halo (e.g., \citealt{Crain13,Anderson16}). Furthermore, only when the dark matter halo is massive enough (e.g., $M_{\rm h}\gtrsim10^{12.7}\rm~M_\odot$ which corresponds to a stellar mass of $M_*\approx10^{11}\rm~M_\odot$ and a rotation velocity of $v_{\rm rot}\approx 300\rm~km~s^{-1}$) could the CGM be gravitationally heated to an X-ray emitting temperature above the peak of the radiative cooling curve (e.g., \citealt{Keres09,vandeVoort16}). Therefore, the extended hot gas in the outer halo is only expected to be detected in the most massive spiral galaxies.

Most existing X-ray observations are only deep enough to detect the X-ray emission from the inner halo (e.g., \citealt{Strickland04,Tullmann06,Li11,Li13a}). In recent years, direct detections of large-scale galactic corona have been reported in a few cases to 50-70~kpc (or typically $\gtrsim 0.1r_{200}$; \citealt{Anderson11,Anderson16,Dai12,Bogdan13a,Bogdan15}). However, most of these observations, except for \citet{Dai12}'s work on UGC~12591, are on less inclined (or largely face-on) galaxies. These observations of face-on galaxies give a significantly higher X-ray luminosity per stellar mass than less massive galaxies or edge-on galaxies with similar stellar masses, although careful rescaling of the X-ray luminosity based on the radial distribution of hot gas has been attempted \citep{Li14}. We herein introduce a small but complete sample of edge-on massive spiral galaxies which is optimized for a statistically meaningful comparison of the hot gas properties.

\subsection{Sample Selection}\label{subsection:SampleProject}

We developed a sample of \emph{isolated}, \emph{massive}, \emph{spiral} galaxies with \emph{moderate angular size} from 7678 \emph{giant} spiral galaxies in the luminosity class range of LC~I-III (supergiant to normal giant according to NED). Our selection criteria are: (1) maximum gas rotation velocity $v_{\rm maxg}\gtrsim300\rm~km~s^{-1}$ (the inclination-corrected velocity, $v_{\rm rot}$, is listed in Table~\ref{table:sample}); (2) Galactic foreground absorption column density $N_{\rm H}<10^{21}\rm~cm^{-2}$; (3) distance $<100\rm~Mpc$; (4) stellar mass $M_*\gtrsim2\times10^{11}\rm~M_\odot$ ($\log M_*>11.3$); (5) $r_{200}<35^\prime$. For galaxies passing these criteria, we finalize our selection by checking their optical images and select only those in the field and with no bright companions within $10^\prime$ ($200\rm~kpc$ at a distance of 70~Mpc). Salient parameters of the selected galaxies are summarized in Table~\ref{table:sample}.


\begin{table*}
\vspace{-0.in}
\begin{center}
\caption{Parameters of the CGM-MASS Sample Galaxies.} 
\footnotesize
\vspace{-0.0in}
\tabcolsep=3.2pt%
\begin{tabular}{lcccccccccccccc}
\hline
Name        & Dist     & Size & $i$ & $N_{\rm H}$ & $v_{\rm maxg}$ & $v_{\rm rot}$ & $m_{\rm K}$ & $\log M_*$ & SFR & $\log M_{\rm 200}$ & $r_{\rm 200}$ & ObsID & $t_{\rm X}$ \\
            & Mpc      & $a\times b$ & deg & $10^{20}\rm cm^{-2}$& $\rm km/s$ & $\rm km/s$ & mag & $\log M_\odot$ & $\rm M_\odot~yr^{-1}$ & $\log M_\odot$ & kpc (arcm) & & ks \\
\hline
NGC 669     & 77.8     & $2.3^\prime\times0.6^\prime$ & 90 & 5.04  & 356.1 & 356.1 & 8.833 & 11.58 & 4.4  & 12.94 & 428 (18.9) & 0741300201 & 123.9 \\
ESO142-G019 & 64.6$^*$ & $2.5^\prime\times1.1^\prime$ & 70.28 & 5.48  & 330.7 & 351.3 & 8.905 & 11.39 & 0.7  & 12.92 & 422 (22.5) & 0741300301 & 91.9  \\
NGC 5908    & 51.9     & $3.4^\prime\times1.6^\prime$ & 65.29 & 1.43  & 315.6 & 347.5 & 8.286 & 11.45 & 8.9  & 12.91 & 417 (27.6) & 0741300101 & 45.5  \\
UGCA 145    & 69.3     & $3.1^\prime\times0.5^\prime$ & 90 & 8.71  & 329.1 & 329.1 & 8.961 & 11.43 & 5.7  & 12.83 & 393 (19.5) & 0741300401 & 111.6 \\
NGC 550     & 93.1     & $1.8^\prime\times0.7^\prime$ & 77.67 & 3.01  & 310.6 & 317.9 & 9.277 & 11.56 & -    & 12.78 & 379 (14.0) & 0741300501 & 73.0 \\
     &      &   &   &   &  &  &  &  &    &  &  & 0741300601 & 75.0 \\
\hline
\end{tabular}\label{table:sample}
\end{center}
\vspace{-0.1in}
Distances without ``*'' are redshift-independent. Diameter of major ($a$) and minor axes ($b$) (described in $D_{\rm 25}$) and the inclination angle ($i$) are obtained from HyperLeda. The foreground absorption column density ($N_{\rm H}$) is obtained from HEASARC web tools. K-band magnitude ($m_{\rm K}$) is obtained from the \emph{2MASS} extended source catalog \citep{Skrutskie06} and $M_*$ is estimated from $m_{\rm K}$ and a color-dependent mass-to-light ratio \citep{Bell01}. The SFR is estimated from the \emph{IRAS} total IR luminosity ($12-100\rm~\mu m$) using \citet{Kennicutt98}'s relation, which could be slightly biased from some recent calibrations \citep{Li16}. The halo mass (with density 200 times of the critical density of the Universe) $M_{\rm 200}$ and virial radius $r_{\rm 200}$ are estimated from the inclination corrected maximum gas rotation velocity $v_{\rm rot}$ (the apparent maximum gas rotation velocity is $v_{\rm maxg}$, both from HyperLeda) with the NFW model (\citealt{Navarro97}). $t_{\rm X}$ is the total duration of the \emph{XMM-Newton} observations.\\
\end{table*}

The above sample selection is optimized for our scientific goals of studying the extended hot galactic corona. Criterion (1) and (4) ensure that the selected galaxies are all massive. The low foreground extinction and distance limit make the galaxies optimized for X-ray observations. The angular size criteria ensures that a large fraction of the extended hot halo ($r_{\rm 200}$) could be covered by a single exposure of \emph{XMM-Newton} (FOV$\approx30^\prime$). The selection criteria on $v_{\rm maxg}$ tends to exclude face-on galaxies. All the sample galaxies have inclination angles $>65^\circ$. The highly inclined orientation could help to reduce the potential contamination from bright X-ray sources in the galactic disk, such as supernova remnants (SNRs), \ion{H}{2} regions, young stellar sources, and HMXBs. These massive galaxies also have relatively weak star formation activity. Adopting a typical definition of starburst galaxies, such as ${\rm SFR}/M_*\gtrsim1.0\rm~M_\odot~yr^{-1}/(10^{10}~M_\odot)$ (e.g., \citealt{Li13a}), all the sample galaxies could be classified as non-starburst.

All the five galaxies in this mini-sample have been observed by \emph{XMM-Newton} in AO-13 and AO-14, with a total exposure time of $\approx0.5\rm~Ms$ (the program has been approved in AO-13; PI: Jiang-Tao Li).

\section{Analysis of the XMM-Newton Data of NGC~5908}\label{section:CaseStudyN5908}

In the present paper, we perform a case study of NGC~5908. Details of the \emph{XMM-Newton} data reduction processes are presented in the appendix (\S\ref{subsection:DataReduction}), including basic data calibration (\S\ref{subsubsection:DataCalibration}), point source detection and subtraction (\S\ref{subsubsection:PointSource}), background analysis (\S\ref{subsubsection:BackAnalysis}), and spectra extraction (\S\ref{subsubsection:SpecExtraction}). Identification and analysis of the X-ray bright point-like sources and the prominent extended soft X-ray features in the FOV are also presented in the appendix (\S\ref{section:PointSrcExtendFeature}). Below we focus on the analysis of the hot corona around NGC~5908.

\subsection{NGC~5908}\label{subsection:N5908}

NGC~5908 is the nearest galaxy in our \emph{XMM-Newton} large program (Table~\ref{table:sample}, $1^\prime=15.1\rm~kpc$). It is an SAb galaxy with a moderately sized bulge. The integrated (in 750-1500~MHz) self-absorption corrected \ion{H}{1} 21-cm line flux is $43.96\rm~Jy~km~s^{-1}$ \citep{Springob05}, corresponding to an atomic gas mass of $M_{\rm HI}=2.8\times10^{10}\rm~M_\odot$ at a distance of 51.9~Mpc \citep{Theureau07}. The specific SFR, estimated from the SFR, the stellar mass $M_*$, and the diameter of the major axis listed in Table~\ref{table:sample}, is just ${\rm SFR}/M_*=0.32\rm~M_\odot~yr^{-1}/(10^{10}~M_\odot)$ or $I_{\rm SFR}=4.3\times10^{-3}\rm~M_\odot~yr^{-1}~kpc^{-2}$, both about twice of the values of the Milky Way \citep{Robitaille10,McMillan11}. For comparison, starburst galaxies typically have ${\rm SFR}/M_*\gtrsim1.0\rm~M_\odot~yr^{-1}/(10^{10}~M_\odot)$ or $I_{\rm SFR}\gtrsim10^{-2}\rm~M_\odot~yr^{-1}~kpc^{-2}$ (e.g., \citealt{Li13a,Li13b,Wang16}). Therefore, this low SFR of NGC~5908 minimizes the stellar feedback effect on the hot corona, especially at large radii. NGC~5908 has a face-on companion galaxy NGC~5905 (SBb, $M_*=4.8\times10^{10}\rm~M_\odot$, ${\rm SFR}=4.6\rm~M_\odot~yr^{-1}$, $v_{\rm rot}=258\rm~km~s^{-1}$, $i=37.49^\circ$), located $\approx200\rm~kpc$ to the northwest in projection. However, as shown in Fig.~\ref{fig:images} and presented in the appendix \S\ref{subsection:ExtendedFeature}, there is no strong evidence of a dense ICM which could significantly affect the extended galactic corona. 

\begin{figure*}
\begin{center}
\epsfig{figure=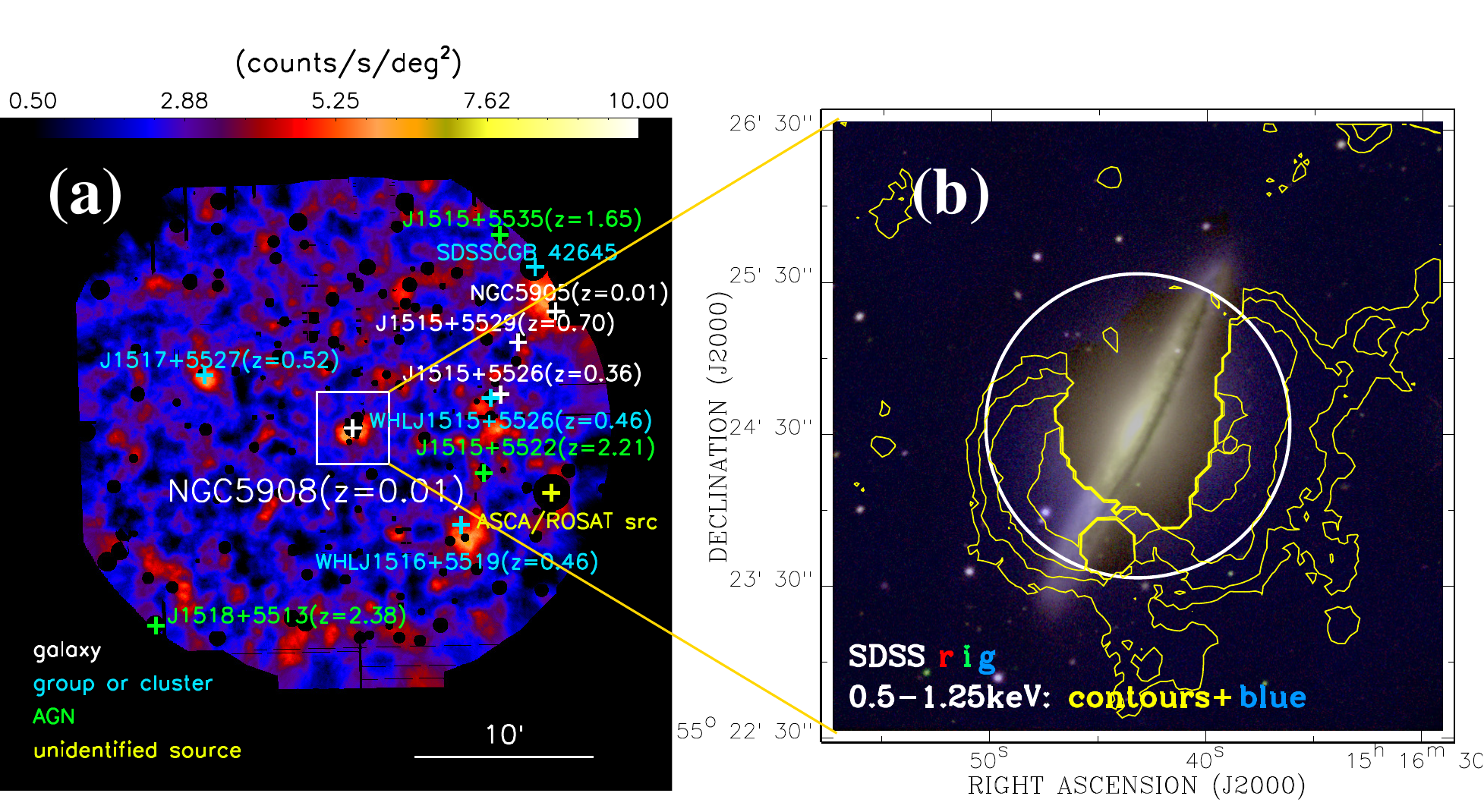,width=1.0\textwidth,angle=0, clip=}
\caption{(a) Point source removed, soft proton and quiescent particle background subtracted, exposure corrected, and adaptively smoothed 0.5-1.25~keV \emph{XMM-Newton} EPIC (MOS-1+MOS-2+PN) image of NGC~5908 and the surrounding area. The color bar in $\rm counts~s^{-1}~deg^{-2}$ is marked on top of the image. The exposure maps of different instruments are scaled to MOS-2 with medium filter before the mosaicing. Some sources are marked with pluses, including an unidentified X-ray bright source previously detected with \emph{ASCA} and \emph{ROSAT}. Different types of sources are marked with different colors, as denoted in the lower left corner. The redshifts of the sources, if found in the literature, are also denoted after their names. The white box marks the central $4^\prime\times4^\prime$ region of NGC~5908, which is further zoomed in in panel~(b). (b) Diffuse soft X-ray emission in the close vicinity of NGC~5908 overlaid on the SDSS tri-color images (r-, i-, g-bands). Yellow contours and the blue color are the same image as shown in panel (a). The contours are at 2, 3, 5, 10~$\sigma$ levels above the background. X-ray bright point sources are masked. The white solid circle with a radius of $1^\prime$ is the region used for spectral analysis as shown in Fig.~\ref{fig:spec}.}\label{fig:images}
\end{center}
\vspace{0.1in}
\end{figure*}

\subsection{Spatial Analysis}\label{subsection:SpatialAnalysis}

As shown in Fig.~\ref{fig:images}b, the diffuse soft X-ray emission extends outside the stellar disk and bulge of NGC~5908, forming a galactic corona. This emission is consistent with being symmetric (except for the removed point sources) at $\sim3\rm~\sigma$ level; there are some extened filamentary structures at $\sim2\rm~\sigma$, which, however, may be artificial caused by the smoothing of the image.

In order to study the distribution of hot gas around NGC~5908, we extract a radial intensity profile in 0.5-1.25~keV from the original calibrated but unsmoothed images. As will be presented in \S\ref{subsection:SpecAnalysis}, the X-ray emission in this band has the largest contribution from hot gas.
The point sources and prominent extended X-ray features are excluded using the masks constructed in \S\ref{subsubsection:PointSource} (Fig.~\ref{fig:masks}), except for the extended emission within $2^\prime$ from the center of NGC~5908.
There are several bright point-like sources located close to the nucleus of NGC~5908 (\S\ref{subsection:PointSrc}). We then extract the profile from $r>0.5^\prime$ and adaptively regroup it to $\rm S/N>7$ for each bin, where the noise includes the contributions from the removed quiescent particle background (QPB).

Stellar sources make a non-negligible contribution in soft X-rays. There are mainly two components of stellar sources related to the old stellar population remaining in the unresolved soft X-ray emission: the LMXBs below the detection limit of $\approx1.9\times10^{38}\rm~ergs~s^{-1}$ (\S\ref{subsubsection:PointSource}) and the collected contribution from individually faint stellar sources such as CVs and ABs (e.g., \citealt{Revnivtsev08}). The contributions from both the unresolved LMXB and CV+AB components can be scaled to the stellar mass of the galaxy. 

We quantitatively estimate the stellar source contribution using the approach adopted in similar X-ray study of some early-type spiral galaxies (e.g., \citealt{Li09,Li11,Li15a}). We first extract a near-IR intensity profile from a calibrated \emph{2MASS} K-band image, after removing the nucleus and some other IR-bright foreground or background sources. We convert the K-band luminosity to stellar mass, using a color-dependent mass-to-light ratio from \citet{Bell01} and the average extinction-corrected B-V color of NGC~5908 (0.815~mag, from the HyperLeda database). The total X-ray luminosity of LMXB per stellar mass can be estimated from their luminosity function (e.g., \citealt{Gilfanov04}). On the other hand, the total X-ray contribution of the CV+AB component can be estimated from deep X-ray observations of nearby low mass elliptical galaxies (e.g., \citealt{Revnivtsev08}). These galaxies have little hot gas, and most of their bright XRBs can be individually detected and removed. The remaining emission is therefore mostly comprised of individually faint stellar sources such as CVs and ABs.

We adopt an X-ray spectral model of LMXB from \citet{Irwin03} (a $\Gamma=1.6$ power law) and a model of CV+AB from \citet{Revnivtsev08} (a 0.5~keV thermal plasma plus a $\Gamma=1.9$ power law). Using these spectral models, we finally convert the K-band intensity profile to the expected 0.5-1.25~keV intensities from LMXBs and CV+AB, respectively. 

We do not add any contribution of young stellar sources to the soft X-ray intensity profile. These sources are expected to follow the spatial distribution of star formation regions, thus are expected to be mostly embedded in the galactic disk. The gaseous disk, as traced by the dustlane (Fig.~\ref{fig:images}b), should have already absorbed most of the soft X-ray emission from these young stellar sources. As a result, we do not detect any significant enhancement of soft X-ray emission along the galactic disk, except for the masked nuclear region (Fig.~\ref{fig:images}b). We will further discuss the stellar sources in the nuclear region in \S\ref{subsection:XrayNuclear} and the residual emission from young stellar sources in the diffuse soft X-ray emission in \S\ref{subsection:HMXBs}.

\begin{figure}
\begin{center}
\epsfig{figure=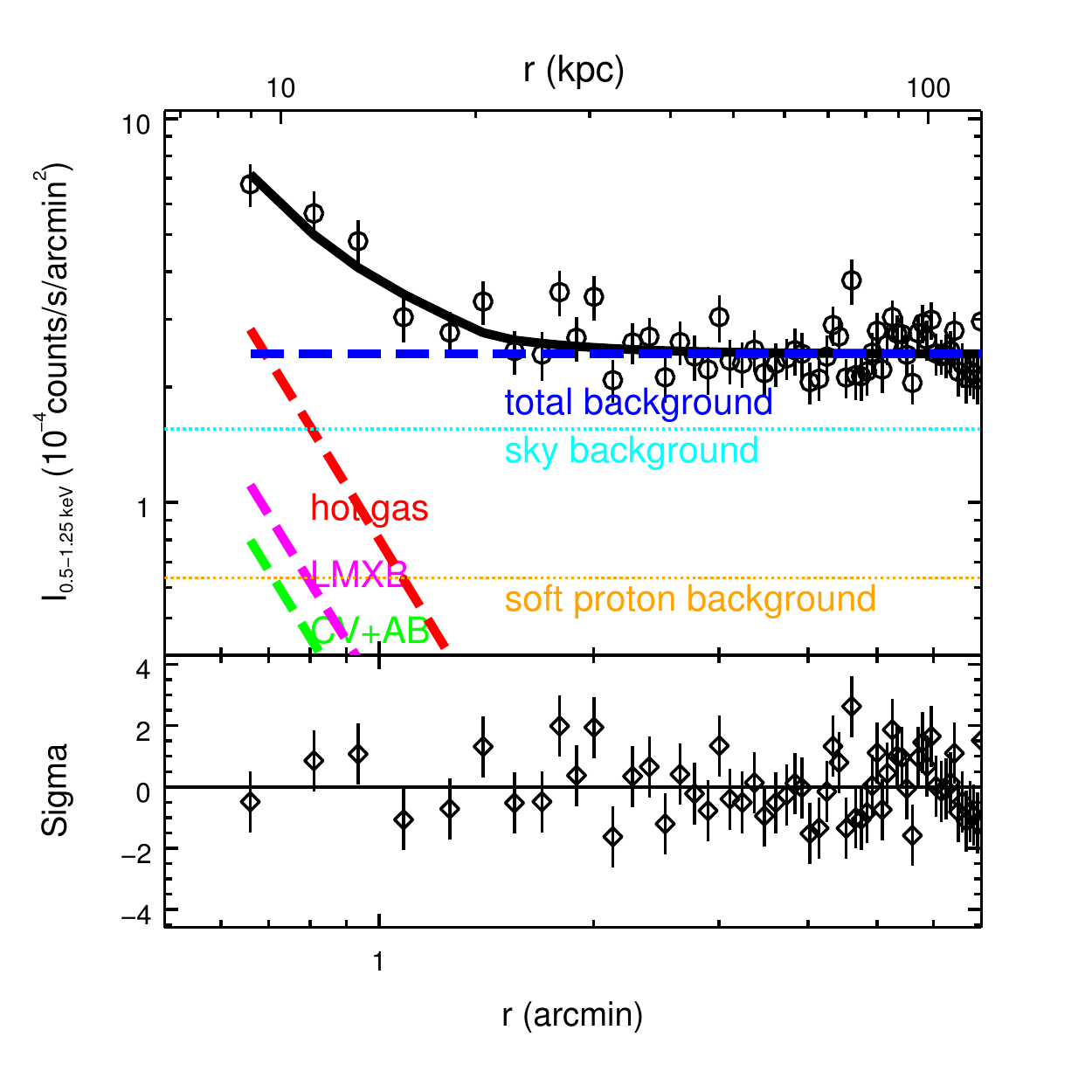,width=0.5\textwidth,angle=0, clip=}
\vspace{-0.3in}
\caption{0.5-1.25~keV radial intensity profile of NGC~5908, created after subtracting the QPB and exposure correction. An X-ray bright nucleus has been masked off (Fig.~\ref{fig:images}b), so the profile starts at $r\approx0.6^\prime$. The best-fit model, shown as a thick black solid curve, is comprised of several components: the sky+SP background (blue dashed), the LMXB and CV+AB contributions estimated from the K-band intensity profile (magenta and green dashed), and a $\beta$-function representing the hot gas distribution (red dashed). For comparison, we also plot the sky (cyan) and SP (orange) background levels with thin dotted lines, separately.}\label{fig:profile}
\end{center}
\end{figure}

We fit the soft X-ray intensity profile with a $\beta$-model ($I_{\rm X}=I_{\rm X,0}[1+(r/r_{\rm core})^2]^{0.5-3\beta}$, after subtracting the stellar (LMXB, CV+AB) contributions, as well as the QPB, soft proton (SP), and sky backgrounds (Fig.~\ref{fig:profile}). Because the nuclear region has been filtered out, the profile can be extracted only at $r\gtrsim0.5^\prime$ (the first data point in Fig.~\ref{fig:profile} is at $r\approx0.6^\prime$). As a result, $r_{\rm core}$ cannot be well constrained. We therefore fix $r_{\rm core}$ at a sufficiently small value of $0.1^\prime$ ($1.5\rm~kpc$), because there is no feature in the radial intensity profile indicating a core radius larger than the radius of the innermost data point. Similar unresolved core in the other giant spirals has been suggested as well (e.g., \citealt{Anderson11,Dai12}). The only free parameters are then $\beta$ and $I_{\rm X,0}$. The best-fit results are $\beta=0.68_{-0.11}^{+0.14}$ and $I_{\rm X,0}=0.097_{-0.073}^{+0.439}\rm~counts~s^{-1}~arcmin^{-2}$ (errors are quoted at the 1~$\sigma$ confidence level).

\subsection{Spectral Analysis}\label{subsection:SpecAnalysis}

We extract spectra from a circular region centered at the nucleus of NGC~5908 and of a radius $r=1^\prime$ (with the nuclear region masked). This region encloses most of the soft X-ray emitting features more than 5~$\sigma$ above the background (Fig.~\ref{fig:images}b). Outside this radius, the diffuse X-ray emission is weak, compared to the multiple stellar and background components (Fig.~\ref{fig:profile}). 

\begin{figure}
\begin{center}
\hspace{-0.3in}
\epsfig{figure=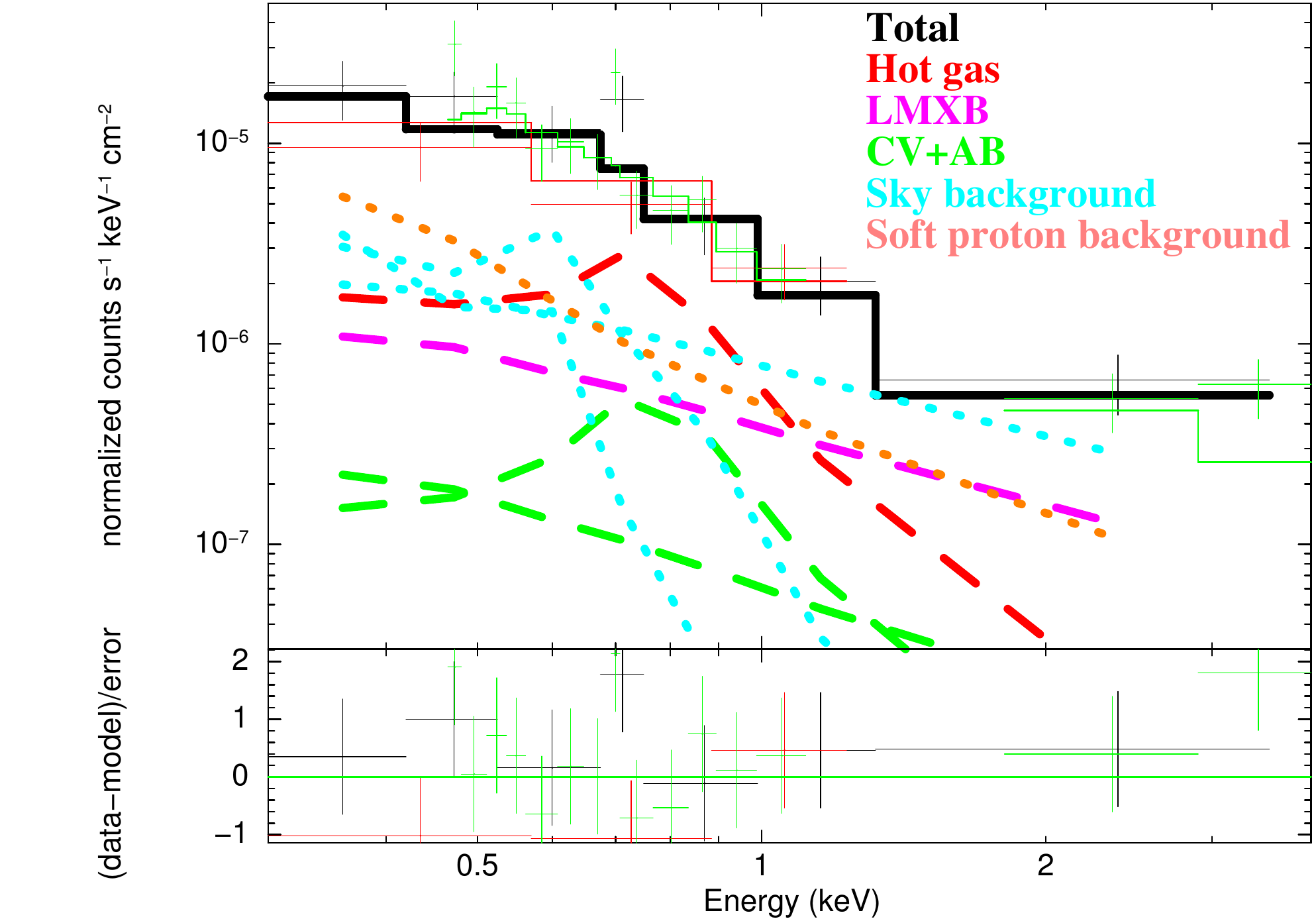,width=0.5\textwidth,angle=0, clip=}
\caption{Spectra of the inner halo ($r<1^\prime$), regrouped to $\rm S/N>3$. Curves representing different model components are the same as those in the radial intensity profile (Fig.~\ref{fig:profile}), and are also denoted on top right. The plotted data points have been scaled with the effective area of each instrument (MOS-1: black; MOS-2: red; PN: green).}\label{fig:spec}
\end{center}
\vspace{0.1in}
\end{figure}

We rescale all the background model components, as well as the LMXB and the CV+AB components (\S\ref{subsection:SpatialAnalysis}), and add them to the model source spectra. All the parameters of the background model components are fixed, which are good enough to describe the background at the location of the source features. 

We use an ``APEC'' model \citep{Smith01} to describe the hot gas emission from the spectral analysis region. The absorption column density is fixed at the Galactic foreground value of $N_{\rm H}=1.43\times10^{20}\rm~cm^{-2}$. We also add a gain correction to the response file of the PN spectrum (``GAIN'' model in XSpec), in order to account for the deficiency in the low-energy calibration of the PN camera \citep{Dennerl04}. Similar as in \citet{Li15b}, the slope of the GAIN is fixed at 1 and the offset is set free.

We first fit the spectra with the temperature ($kT$), abundance ($Z$), and normalization of the APEC component all allowed to freely vary. Therefore, there are four free parameters, including the offset of GAIN. It is obvious that these parameters cannot be well constrained simultaneously, because the counting statistic is poor, the spectral decomposition is complicated, and the spectra are nearly featureless in soft X-ray (Fig.~\ref{fig:spec}). 

The featureless spectra indicate that the metal abundance of hot gas may be significantly subsolar, because a higher metallicity should result in detectable features such as the Fe L-shell bumps at $\sim1\rm~keV$ (e.g., \citealt{Li09,Li15a}). Because the metallicity of hot gas is poorly constrained, we fix it at 0.2~solar, following the direct X-ray constraint on some similar massive spiral galaxies (e.g., \citealt{Bogdan15,Anderson16}). 

The best-fit temperature and 0.5-2~keV luminosity of the hot gas within the spectral extraction region are $kT=0.38_{-0.09}^{+0.64}\rm~keV$ and $L_{\rm X}=4.6_{-1.5}^{+1.8}\times10^{38}\rm~ergs~s^{-1}$ (errors are at the 1~$\sigma$ level).

\subsection{X-ray emission from the nuclear region}\label{subsection:XrayNuclear}


In this section, we discuss the X-ray emission from the nuclear region of NGC~5908. As shown in Fig.~\ref{fig:pointsrcimg}b, in addition to the AGN, another X-ray bright point-like source is detected to the northwest of the nucleus and is apparently located in the galactic disk. This source probably has an optically bright (but faint in K-band) counterpart located $\approx4^{\prime\prime}$ away. Its 0.3-7.2~keV counts rate is $\lesssim40\%$ of the AGN. 

Between the AGN and this source is an apparently diffuse X-ray feature, which may however represent a collection of young stellar sources in the galactic disk. \citet{Sazonov06} found a positive correlation between the luminosity and the 2-10~keV/0.1-2.4~keV luminosity ratio of ABs and young stars. Because individual young stars typically have 0.1-2.4~keV luminoisty $<10^{30}\rm~ergs~s^{-1}$ while ABs in \citet{Sazonov06}'s sample are more X-ray luminous, we expect the young stars should have a softer spectrum than ABs so an even less contribution after absorbed by the almost edge-on gaseous disk. On the other hand, the contribution from HMXBs may not be negligible. Using the $L_{\rm 0.5-8keV,HMXB}-{\rm SFR}$ relation from \citet{Mineo12}, we can estimate the total 0.5-8~keV X-ray luminosity of HMXBs in NGC~5908 by adopting the SFR listed in Table~\ref{table:sample}. This luminosity is $2.3\times10^{40}\rm~ergs~s^{-1}$. In contrast, the total 0.5-8~keV X-ray luminosity of the non-AGN sources in the nuclear region (the point-like source northwest to the nucleus and the apparently diffuse X-ray features) is $1.9\times10^{40}\rm~ergs~s^{-1}$. Considering the scatter in the $L_{\rm 0.5-8keV,HMXB}-{\rm SFR}$ relation (e.g., \citealt{Mineo12}) and the uncertainties in estimating the SFR (see \S\ref{subsection:HMXBs}), we conclude that most of the X-ray emission from HMXBs is detected from the nuclear region.

\section{Results and Discussions}\label{section:ResultsDiscussion}

\subsection{Contribution from HMXBs}\label{subsection:HMXBs}

The SFR of NGC~5908 is $8.9\rm~M_\odot~yr^{-1}$ (Table~\ref{table:sample}), indicating a possibly significant soft X-ray contribution of young stellar sources below the point source detection limit of $1.9\times10^{38}\rm~ergs~s^{-1}$. The most important young stellar source in X-ray is HMXB (e.g., \citealt{Mineo12}). We therefore need to estimate the residual HMXB contribution to the measured ``hot gas'' emission.

We first assume the spatial distribution of star formation regions exactly follows the \emph{WISE} 22$\rm~\mu m$ emission. This assumption means that there is no AGN (which is only important at small radii) or the dust heated by the general interstellar radiation field which is not related to current star formation (can be important at large radii; \citealt{Temi09}), and the extinction in mid-IR is negligible (could be significant for edge-on galaxies; \citealt{Li16}). We then calculate the fraction of 22$\rm~\mu m$ emission enclosed by the \emph{XMM-Newton} spectral analysis region (Fig.~\ref{fig:images}b), which is $\approx16\%$ of the total 22$\rm~\mu m$ emission of NGC~5908. We therefore obtain the SFR from the X-ray spectral analysis region to be $1.4\rm~M_\odot~yr^{-1}$.

We also estimate the SFR from the 1.4~GHz radio continuum flux (38.6~mJy; \citealt{Condon02}) and the radio-IR relationship of edge-on galaxies from \citet{Li16}. Assuming a radio spectral index of 1.0, we obtain a SFR of $\approx2.9\rm~M_\odot~yr^{-1}$ for the whole galaxy, compared to $8.9\rm~M_\odot~yr^{-1}$ obtained from IR observations (Table~\ref{table:sample}). It is expected that the SFR estimated this way is 2-3 times lower than that predicted with the radio-IR relationship of face-on galaxies, largely because of the strong extinction even in IR in edge-on cases. Discussions on the radio and IR star formation tracers of edge-on galaxies are beyond the scope of the present paper, and can be found in more details in \citet{Li16}. We just emphasize that the adopted SFR of $1.4\rm~M_\odot~yr^{-1}$ from the X-ray spectral analysis region may be overestimated. The overestimation could be even more significant if we consider the contribution from the cold dust at large radii.

We then estimate the total X-ray luminosity from HMXBs below the detection limit of $1.9\times10^{38}\rm~ergs~s^{-1}$ using the HMXB luminosity function (LF) from \citet{Mineo12}. We also convert the 0.5-8~keV luminosity as adopted in \citet{Mineo12} to 0.5-2~keV as adopted for the hot gas emission in this paper, using an average HMXB spectrum from \citet{Swartz04} (i.e., an $\Gamma=2.0$ power law). The final 0.5-2~keV HMXB luminosity below the detection limit is: $L_{\rm X,HMXB}/{\rm SFR}=2.2\times10^{38}\rm~ergs~s^{-1}/(M_\odot~yr^{-1})$.

Assuming the spatial distribution of unresolved HMXB exactly follows the 22$\rm~\mu m$ emission, we obtain a HMXB contribution to the ``hot gas'' component within the spectral analysis region to be $3.1\times10^{38}\rm~ergs~s^{-1}$. This is $\sim68\%$ of the ``hot gas'' component, so HMXBs could make a significant contribution to the unresolved soft X-ray emission. However, as discussed above, the HMXB contribution may be significantly overestimated, so we still expect there is a considerable fraction of diffuse soft X-ray emission from the true hot gas. In any cases, the properties of the hot gas are largely dependent on the exact amount of young stellar contribution, so could be very uncertain. The hot gas luminosities listed in Table~\ref{table:pararadius}, especially at small radii, can only be adopted as upper limits.

\subsection{Physical Properties of the Corona}\label{subsection:CoronaProperty}


The directly measured X-ray luminosity of hot gas within the spectral analysis region (\S\ref{subsection:SpecAnalysis}) needs to be further corrected for the masked nuclear region. We rescale this luminosity based on the best-fit X-ray intensity profile of the hot gas component (Fig.~\ref{fig:profile}), assuming that the best-fit $\beta$-model can be extrapolated to both smaller and larger radii and there is no young stellar source contribution to the diffuse soft X-ray emission. This rescaled luminosity is listed in the first row of Table~\ref{table:pararadius}. 

\begin{table*}
\vspace{-0.in}
\begin{center}
\caption{Inferred parameters of the diffuse hot gas within various radii} 
\footnotesize
\vspace{-0.0in}
\tabcolsep=4.pt%
\begin{tabular}{lcccccccccccccc}
\hline
            &  $L_{\rm X, 0.5-2keV}$       & $\stretchleftright{\langle}{n_{\rm e}}{\rangle}$  		           & $M_{\rm hot}$                       & $\stretchleftright{\langle}{P_{\rm hot}}{\rangle}$                   & $E_{\rm hot}$                   & $\stretchleftright{\langle}{t_{\rm cool}}{\rangle}$ \\ 
            &  $\rm 10^{39}ergs~s^{-1}$  & $10^{-3}f^{-1/2}\rm cm^{-3}$ & $10^{10}f^{1/2}\rm M_\odot$ & $f^{-1/2}\rm eV~cm^{-3}$ & $10^{57}f^{1/2}\rm ergs$ & $f^{1/2}\rm Gyr$                  \\ 
\hline
$r<15\rm~kpc$ ($1.00^\prime$, $0.036r_{\rm200}$) & $6.83_{-2.20}^{+2.73}$ & $6.16_{-4.36}^{+7.61}$       & $0.231_{-0.163}^{+0.285}$ & $2.00_{-1.08}^{+6.66}$       & $1.36_{-0.73}^{+4.52}$ & $0.97~(<2.13)$ \\
$r<25\rm~kpc$ ($1.66^\prime$, $0.06r_{\rm200}$, $r_{\rm cool}$) & $7.09_{-2.28}^{+2.83}$ & $2.94_{-2.08}^{+3.64}$       & $0.502_{-0.355}^{+0.620}$ & $0.955_{-0.515}^{+3.182}$       & $2.95_{-1.59}^{+9.82}$ & $2.04~(<4.45)$ \\
$r<50\rm~kpc$ ($3.31^\prime$, $0.12r_{\rm200}$)           & $7.28_{-2.34}^{+2.91}$ & $1.05_{-0.75}^{+1.30}$ & $1.44_{-1.02}^{+1.78}$           & $0.342_{-0.184}^{+1.140}$ & $8.44_{-4.55}^{+28.14}$    & $5.69~(<12.42)$ \\
\hline
\end{tabular}\label{table:pararadius}
\end{center}
\vspace{-0.1in}
The hot gas parameters are scaled based on the best-fit X-ray intensity profile of the hot gas component, without subtracting the young stellar contributions. Errors are 1~$\sigma$ and statistical only. Many systematic uncertainties, such as the poorly constrained metallicity and radial intensity profile at large galactocentric radii, are not included in the errors here. The luminosity $L_{\rm X}$, mass ($M_{\rm hot}$), and thermal energy ($E_{\rm hot}$) are the total values, while the electron number density ($n_{\rm e}$), the thermal pressure ($P_{\rm hot}$), and the radiative cooling timescale ($t_{\rm cool}$) are average values.
\end{table*}

We also estimate other parameters of the hot gas based on the spectral fitting result at $r<1^\prime$. Assuming spherical symmetric distribution of the hot gas, the emission measure (EM) could be obtained from the normalization of the APEC model. 
Adopting a metal abundance of $Z=0.2~Z_\odot$ (based on the latest measurement by \citet{Anderson16} for NGC~1961), we could further obtain the number ratio of electron to Hydrogen atom (1.20) and the mean atomic weight per atom (1.27). These parameters do not change significantly with the assumed abundance. The electron number density ($n_{\rm e}$) could thus be estimated from the EM, and depends on the unknown volume filling factor $f$ in $f^{-1/2}$. The total mass of the hot gas ($M_{\rm hot}$) is also estimated based on $n_{\rm e}$ and the assumed spherical symmetric geometry. The best-fit hot gas temperature is $kT=0.38_{-0.09}^{+0.64}\rm~keV$ (\S\ref{subsection:SpecAnalysis}). We further obtain the average thermal pressure ($P_{\rm hot}$) and the total thermal energy ($E_{\rm hot}$) contained in the hot gas, based on $n_{\rm e}$, $kT$, $f$, and the volume of the region. We estimate the radiative cooling timescale $t_{\rm cool}$ as:
\begin{equation}\label{equi:CoolingFunction}
t_{\rm cool}=\frac{1.5nkT}{\Lambda n_{\rm e}(n-n_{\rm e})},
\end{equation}
where $n$ is the total particle number density, and $\Lambda$ is the normalized cooling rate of the plasma assumed in collisionally ionized equilibrium (CIE). Adopting an abundance of $Z/Z_\odot=0.2$, we can obtain $n=1.92n_{\rm e}$. We adopt $\Lambda=10^{-23.00}$ for $kT=10^{6.65}\rm~K$ and $[Fe/H]=-1.0$ CIE plasma from \citet{Sutherland93}. This cooling rate should not be significantly different from the $kT=0.38\rm~keV$, $Z=0.2~Z_\odot$ gas discussed here.


The above parameters are only directly measured from the spectral analysis region with $r<1^\prime$. At larger radii, the counting statistics is too poor for us to directly measure the hot gas properties. We therefore estimate the hot gas parameters based on the spectral model of the inner halo and the best-fit $\beta$-model of the hot gas radial distribution. The average or total values of the hot gas parameters within a given radius in units of $r_{\rm200}$ are rescaled from the innermost spectral analysis region. The results are summarized in Table~\ref{table:pararadius}. Since the contribution from young stellar sources is very uncertain (\S\ref{subsection:HMXBs}), we do not subtract this component when estimating the hot gas properties. As a result, all the parameters listed in Table~\ref{table:pararadius} could be adopted as upper limits, except for the radiative cooling timescale (lower limit in the innermost region, while unaffected at larger radii since there will be little stellar source contribution beyond the stellar disk and bulge).


We further define the cooling radius ($r_{\rm cool}$) as the radius at which $t_{\rm cool}=10\rm~Gyr$, within which the hot gas could radiatively cool and be accreted onto the galaxy. From Eq.~\ref{equi:CoolingFunction}, we obtain $t_{\rm cool}=10\rm~Gyr$ at $n_{\rm e}\approx6\times10^{-4}\rm~cm^{-3}$. Since the soft X-ray intensity $I_{\rm X}\propto n_{\rm e}^2$, we can estimate the $n_{\rm e}$ distribution from the hot gas soft X-ray intensity profile  (Fig.~\ref{fig:profile}), assuming there is no radial variation of the hot gas temperature and metallicity. Adopting the best-fit $\beta$-model of the hot gas component, $r_{\rm cool}$ should be $\approx25\rm~kpc$ ($\approx0.06 r_{\rm 200}$). The cooling radius corresponds to an angular size of $\approx 1.66^\prime$, which is also approximately where we have confidently detected the hot gas emission (Fig.~\ref{fig:profile}). We further estimate the hot gas properties within this radius and list them in the second row of Table~\ref{table:pararadius}.

The total radiative cooling rate of the hot gas within the cooling radius is $\dot{M}_{\rm hot}\sim2.5\rm~M_\odot~yr^{-1}$. $\dot{M}_{\rm hot}$ is estimated using the total hot gas mass and the average radiative cooling timescale listed in Table~\ref{table:pararadius}. Considering the large uncertainties in the estimation of both the hot gas properties and the SFR, we may conclude that the radiative cooling of hot gas could at least significantly contribute in replenishing the gas consumed in star formation.

\subsection{Baryon budget within the cooling radius}\label{subsection:Baryon}

We quantitatively estimate the total baryon budget of this relatively isolated, massive spiral galaxy.
As presented in \S\ref{subsection:N5908}, the cold atomic gas mass is estimated from the integrated self-absorption corrected \ion{H}{1} 21-cm line flux \citep{Springob05}, and is $M_{\rm HI}=2.8\times10^{10}\rm~M_\odot$.
We are not aware of a published mass measurement of molecular gas in NGC~5908. However, we may roughly estimate the molecular gas mass from the SFR in Table~\ref{table:sample} and the recent calibration of the star formation law by \citet{Leroy13}. The molecular gas estimated this way is $M_{\rm H_2}\approx2\times10^{10}\rm~M_\odot$, comparable to $M_{\rm HI}$.
The stellar mass in Table~\ref{table:sample} ($2.8\times10^{11}\rm~M_\odot$) is calculated based on the K-band magnitude of the whole galaxy, without subtracting the contribution from the AGN. We therefore adopt it as an upper limit to the total stellar mass $M_*$. After removing the AGN and a few bright point-like sources (which may be foreground stars), we obtain a stellar mass of $M_*=1.4\times10^{11}\rm~M_\odot$. As NGC~5908 has a small angular size ($3.4^\prime\times1.6^\prime$; Table~\ref{table:sample}), the removal of these sources could significantly affect the photometry in K-band, leading to an underestimate of the stellar mass. We therefore adopt a stellar mass of $M_*=(2.1\pm0.7)\times10^{11}\rm~M_\odot$. The error only indicates the difference in near-IR photometry between our estimate after removing point sources and those from the literature of the whole galaxy.

Assuming the filling factor $f=1$, the total baryon mass within the cooling radius, including the cold atomic and molecular gases, the stars, and the hot gas, is $M_{\rm b}\approx2.63\times10^{11}\rm~M_\odot$. Here we have assumed that all the cold gas and stellar mass are enclosed within the cooling radius, which is comparable to the radius of the optical major axis of the galaxy. Within the cooling radius, the hot gas only accounts for a small fraction of the total baryon content ($\approx1.9\%$). 

\subsection{Comparison to other galaxies}\label{subsection:Compare}

Here we briefly compare the hot halo properties of NGC~5908 to previous X-ray observations of nearby galaxies.

\begin{figure}
\begin{center}
\hspace{-0.2in}
\epsfig{figure=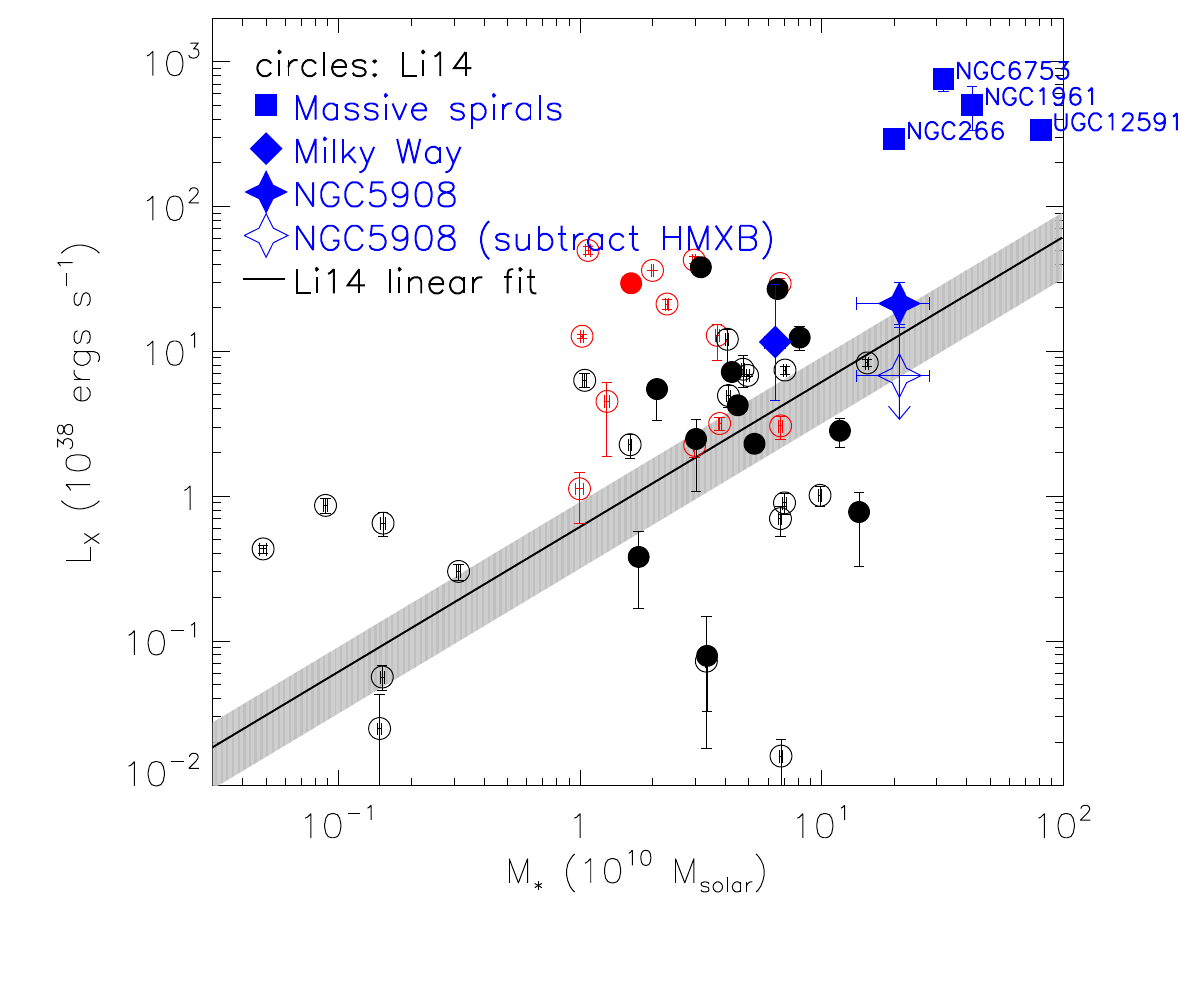,width=0.5\textwidth,angle=0, clip=}
\vspace{-0.3in}
\caption{The 0.5-2~keV X-ray luminosity ($L_{\rm X}$) in $(0.01-0.1)r_{\rm 200}$ v.s. the stellar mass $M_*$ of NGC~5908 in comparison with other galaxies. The X-ray luminosity of the Milky Way Galaxy [scaled to $(0.01-0.1)r_{\rm 200}$, $(1.16_{-0.70}^{+1.74})\times10^{39}\rm~ergs~s^{-1}$] is obtained from \citet{Snowden97}, while the error range is from \citet{Miller15}. The stellar mass of the Milky Way [$(6.43\pm0.63)\times10^{10}\rm~M_\odot$] is obtained from \citet{McMillan11}. The circles represent the nearby edge-on disk galaxies studied in \citet{Li14}, $L_{\rm X}$ of which have also been rescaled to $(0.01-0.1)r_{\rm 200}$. Red/black circles represent starburst/non-starburst galaxies, while open/filled circles are field/clustered galaxies, respectively. The solid line is a linear fit to the non-starburst field galaxies in \citet{Li14} (i.e., the black open circles) and the shaded region represents the 1-$\sigma$ confidence range, reflecting the normalization uncertainty only. For massive spiral galaxies (blue filled squares), NGC~6753 is from \citet{Bogdan13a}; NGC~266 is from \citet{Bogdan13b}; NGC~1961 is from \citet{Anderson11}; and UGC~12591, the other edge-on one similar to NGC~5908, is from \citet{Dai12}.}\label{fig:compare}
\end{center}
\end{figure}


Fig.~\ref{fig:compare} compares the 0.5-2~keV luminosity of hot gas within $(0.01-0.1)r_{\rm 200}$ of NGC~5908 to those of the highly-inclined disk galaxies from \citet{Li14}. We extrapolate the X-ray luminosity measured in the spectral extraction region to $(0.01-0.1)r_{\rm 200}$ ($2.14_{-0.69}^{+0.86}\times10^{39}\rm~ergs~s^{-1}$), using the same method as described in \S\ref{subsection:CoronaProperty}, in order to match the photometry region adopted in \citet{Li14}. We also estimate the truly hot gas emission after subtracting the estimated HMXB contribution (adopting $L_{\rm X,HMXB}=3.1\times10^{38}\rm~ergs~s^{-1}$ for the \emph{XMM-Newton} spectral analysis region; \S\ref{subsection:HMXBs}), which is $0.68 (<1.54)\times10^{39}\rm~ergs~s^{-1}$ after rescaled to $(0.01-0.1)r_{\rm 200}$. However, as discussed in \S\ref{subsection:HMXBs}, we caution that the HMXB contribution may be over-estimated so this luminosity could be adopted as a lower limit.

\citet{Li14}'s sample (using the original sample of \citealt{Li13a,Li13b}) does not include extremely massive spiral galaxies with $M_*\gtrsim2\times10^{11}\rm~M_\odot$. We then include X-ray measurements of massive spiral galaxies from other works for comparison. These galaxies include the Milky Way \citep{Snowden97,Miller15}, UGC~12591 \citep{Dai12}, NGC~1961 \citep{Anderson11,Anderson16,Bogdan13a}, NGC~6753 \citep{Bogdan13a}, and NGC~266 \citep{Bogdan13b}. We caution that most of the massive spiral galaxies from previous works, except for the Milky Way and UGC~12591, have low inclination angles ($\lesssim45^\circ$). Although we have already rescaled the X-ray luminosity of all the galaxies to 0.5-2~keV within $(0.01-0.1)r_{\rm 200}$, there may still be some biases in data homogenization in these nearly face-on cases.

Our X-ray measurement of NGC~5908 indicates that there is a large scatter of X-ray emission from hot gas for galaxies with $M_*\gtrsim2\times10^{11}\rm~M_\odot$. This is also the mass range where the linearity of stellar mass-halo mass relation significantly breaks down (the relation steepens with the mass of galaxies, e.g., \citealt{Behroozi10}). There are currently too few such massive spiral galaxies (especially edge-on ones) with $M_*\gtrsim2\times10^{11}\rm~M_\odot$ to argue whether or not there is a steeper $L_{\rm X}-M_*$ relationship above a certain threshold of $M_*$ and how important the other galaxy properties (e.g., local environment and stellar mass assembly history) may be in determining the X-ray luminosity. We fit the $L_{\rm X}-M_*$ relationship for non-starburst field galaxies from \citet{Li14} as a fiducial relation to compare with. Starburst or clustered galaxies may have hot coronae produced in other ways than the hot mode accretion expected in massive spiral galaxies. The $L_{\rm X}-M_*$ relationship for such a ``clean'' sample (black open circles in Fig.~\ref{fig:compare}) can be characterized with a linear relation, but the data used to fit this relation shows a large scatter of $rms\approx0.9\rm~dex$. NGC~5908 is consistent with such a linear relation defined with lower-mass non-starburst spiral galaxies within $\approx1~\sigma$. On the other hand, UGC~12591, which is clearly more massive, shows a significant departure (at $3.8~\sigma$) from the linear $L_{\rm X}-M_*$ relationship. In comparison, the three face-on galaxies are systematically more X-ray luminous than edge-on and lower mass galaxies. Some of these X-ray luminous face-on galaxies, however, also tend to have higher SFR ($>10\rm~M_\odot~yr^{-1}$ for NGC~1961 and NGC~6753). It is therefore not clear what is the exact reason for their higher X-ray luminosities.

\section{Summary and Conclusions}\label{section:Summary}

The CGM-MASS project is a systematic study of the multi-phase CGM around extremely massive isolated spiral galaxies in the local Universe. The \emph{XMM-Newton} observations of the sample galaxies have been taken to explore the role of the extended hot halo in galaxy evolution. In this paper, we have analyzed the \emph{XMM-Newton} data of NGC~5908, as the first detailed case study of the CGM-MASS project. Although most of the results and discussions will be presented in follow-up papers, we herein summarize the results based on this case study.

(1) After removing the X-ray bright nucleus, the residual emission of excised bright point sources, the stellar contributions of faint LMXBs and CVs+ABs, as well as the multiple background components, the remaining diffuse X-ray emission of NGC~5908 is significantly more extended than the stellar light. The best-fit $\beta$-model to the hot gas component of the 0.5-1.25~keV radial intensity profile has a slope of $\beta=0.68_{-0.11}^{+0.14}$. The diffuse soft X-ray emission is detected out to $\approx2^\prime$, or $\approx30\rm~kpc$ from the nucleus, extends significantly beyond the stellar disk in the vertical direction. 

(2) The total unresolved 0.5-2~keV luminosity within $r=1^\prime=15\rm~kpc$, after removing the contributions from bright point sources and old stellar populations, is $L_{\rm X}=6.8_{-2.2}^{+2.7}\times10^{39}\rm~ergs~s^{-1}$. The contributions from young stellar population are not well constrained, but a rough estimate indicates that the HMXB could contribute up to $\sim68\%$ of the unresolved soft X-ray emission. We further estimate physical parameters of the hot gas within various radii based on the extrapolation of the best-fit hot gas intensity profile. These hot gas parameters include the average electron number density $n_{\rm e}$, thermal pressure $P_{\rm hot}$, radiative cooling timescale $t_{\rm cool}$, as well as the total mass $M_{\rm hot}$ and thermal energy $E_{\rm hot}$. In particular, the cooling radius within which the hot gas could cool radiatively within 10~Gyr is $r_{\rm cool}\approx25\rm~kpc$, comparable to the outermost radius where the hot gas emission is directly detected. Within this cooling radius, the total radiative cooling rate of the hot gas is $\dot{M}_{\rm hot}\sim2.5\rm~M_\odot~yr^{-1}$. Considering the large uncertainties in the estimation of both the hot gas properties and the SFR, this rate indicates that the radiative cooling of hot gas could significantly contribute in replenishing the gas consumed in star formation.

(3) We estimate the baryon budget of NGC~5908. Adding the mass of cold atomic and molecular gases, hot gas, and stars, the total baryon mass within the cooling radius is $M_{\rm b}\approx2.63\times10^{11}\rm~M_\odot$, dominated by the stellar mass. The hot gas only accounts for $\approx1.9\%$ of the total baryon content within the cooling radius, or where the X-ray emission is directly detected.

(4) It is expected that massive spiral galaxies have a larger fraction of their hot coronae produced by the accretion of external gas, so appear to be more X-ray luminous than lower mass galaxies. We do not find such a trend up to the mass of NGC~5908. Our \emph{XMM-Newton} observation indicates that NGC~5908 is consistent with lower mass non-starburst spiral galaxies on the linear $L_{\rm X}-M_*$ relation and is significantly less luminous than some face-on spiral galaxies with similar stellar mass.

\bigskip
\noindent\textbf{\uppercase{acknowledgements}}
\smallskip\\
\noindent The authors would like to acknowledge the \emph{XMM-Newton} helpdesk for helpful discussions on data reduction and Edmund Hodges-Kluck on his helpful comments on the paper. The authors also acknowledge the anounymous referee for his or her constructive comments and suggestions. JTL acknowledges the financial support from NASA through the grants NNX13AE87G, NNH14ZDA001N, and NNX15AM93G. QDW is supported by NASA via a subcontract of the grant NNX15AM93G. RAC is a Royal Society University Research Fellow.

\appendix

\section{XMM-Newton Data Reduction}\label{subsection:DataReduction}

\subsection{Data Calibration}\label{subsubsection:DataCalibration}

\begin{figure*}
\begin{center}
\epsfig{figure=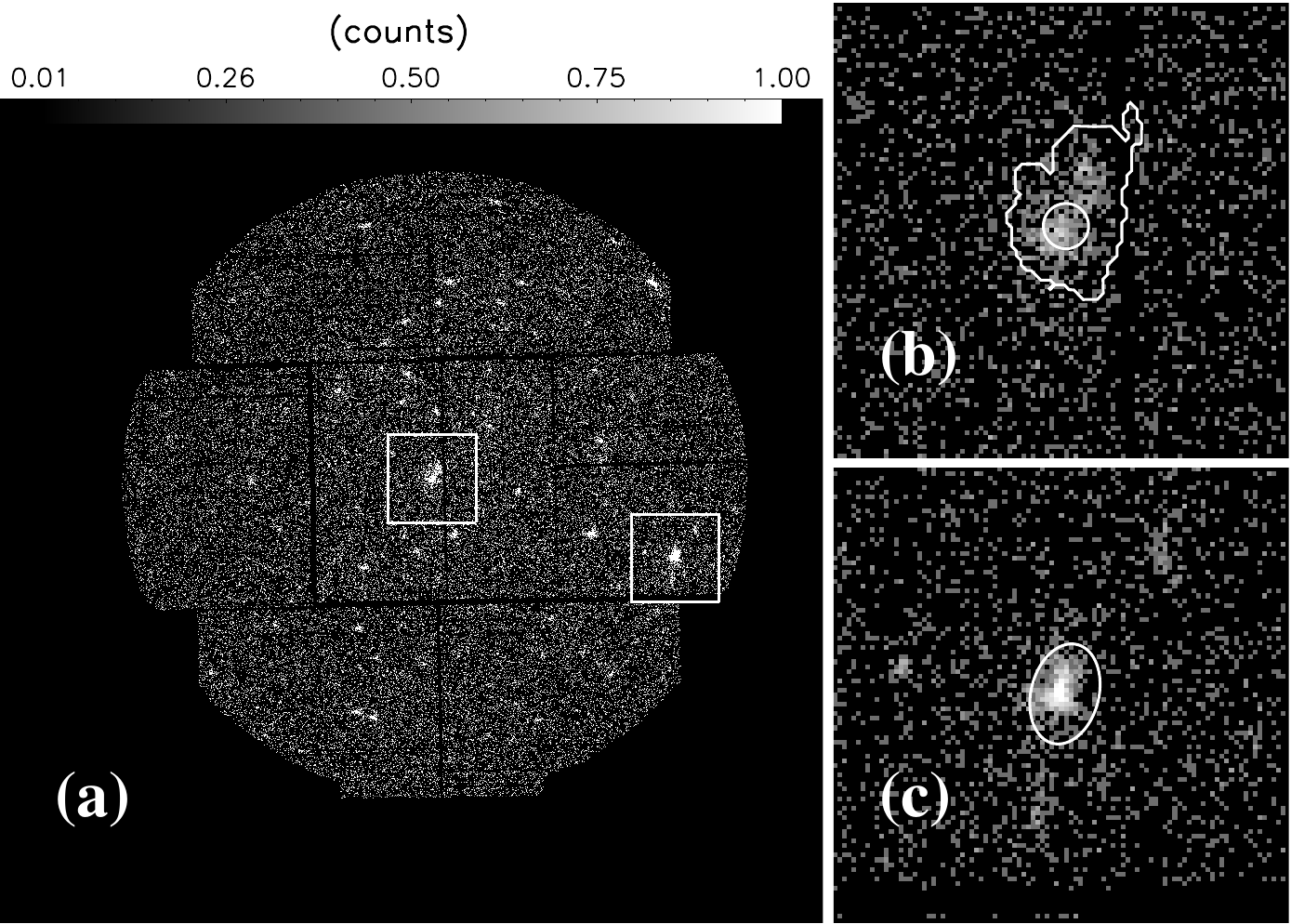,width=1.0\textwidth,angle=0, clip=}
\caption{(a) Calibrated MOS-2 counts image of NGC~5908. The two $4^\prime\times4^\prime$ regions are centered at the nucleus of NGC~5908 and the brightest X-ray point source previously detected by \emph{ASCA/ROSAT}, repectively. (b,c) Zoomed-in MOS-2 counts images of the two boxes shown in panels~(a,b). The circle and ellipse in the two panels are the regions used to extracted spectra of the point sources (Fig.~\ref{fig:pointsrcspec}). The polygon region in panel~(b) is used to remove the nuclear point sources when extracting the diffuse X-ray radial intensity profile (Fig.~\ref{fig:profile}) and spectra (Fig.~\ref{fig:spec}).}\label{fig:pointsrcimg}
\end{center}
\end{figure*}

We calibrate the \emph{XMM-Newton} data of NGC~5908 (Table~\ref{table:sample}) based on the \emph{XMM-Newton} Science Analysis Software (SAS) v14.0.0, with basic steps similar to those described in \citet{Li15b}. 
The Observation Data Files (ODF) for each of the EPIC instruments (MOS-1, MOS-2, and PN) are reprocessed using the SAS tasks \emph{emchain} and \emph{epchain}. 
We identify and tag the low-energy noise in the MOS CCDs in anomalous states, i.e., with an elevated event rates between 0 and 1~keV, using the SAS task \emph{emtaglenoise}, following the algorithm described in \citet{Kuntz08}. All events in all CCDs tagged as noisy are then filtered out. 
We further run \emph{mos-filter} and \emph{pn-filter} on MOS and PN data, respectively. These tools identify good time intervals (GTIs) by screening for periods of soft proton flaring and also adopt the standard parameters for \emph{PATTERN}. The GTI analysis indicates weak background flare during the observation. 
We then remove events belonging to ``wrong'' or ``suspicious'' pixels/rows and/or frames by adopting some additional canned screening set of \emph{FLAG} values ($\#XMMEA\_EM$ for MOS and $FLAG==0$ for PN). We keep the events in the corner of the FOV in the screening processes.
The resultant effective exposure times for the ``cleaned'' event files are 41.35, 42.01, 33.07~ks for MOS-1, MOS-2, and PN, respectively, or about 93.7\%, 95.2\%, and 82.2\% of the total exposure time. Out-of-Time (OoT) events have a non-negligible impact on the PN data and are also reprocessed with the SAS task \emph{epchain} and filtered in an identical way to the primary PN datasets. This OoT events are further subtracted in the following imaging, spatial, and spectral analyses. An example of the calibrated event image of MOS-2 is shown in Fig.~\ref{fig:pointsrcimg}a.

The instrument (telescope+detector) response is not flat, i.e., the effective area at a given energy depends on the position in the focal plane. This vignetting effect is extremely important for the analysis of inhomogeneous extended source such as the diffuse X-ray emission from galactic halos. We apply the SAS task \emph{evigweight} to both the OoT and the primary events lists, weighting each EPIC events with inverse effective area over one exposure, so that the derived event list is equivalent to what one would get for a flat instrument. This allows the use of single on-axis Ancillary Response Files (ARF) for each instrument, and does not require further effective area corrections in the following spectral analyses.

\subsection{Point Source Detection}\label{subsubsection:PointSource}

\begin{figure*}
\begin{center}
\epsfig{figure=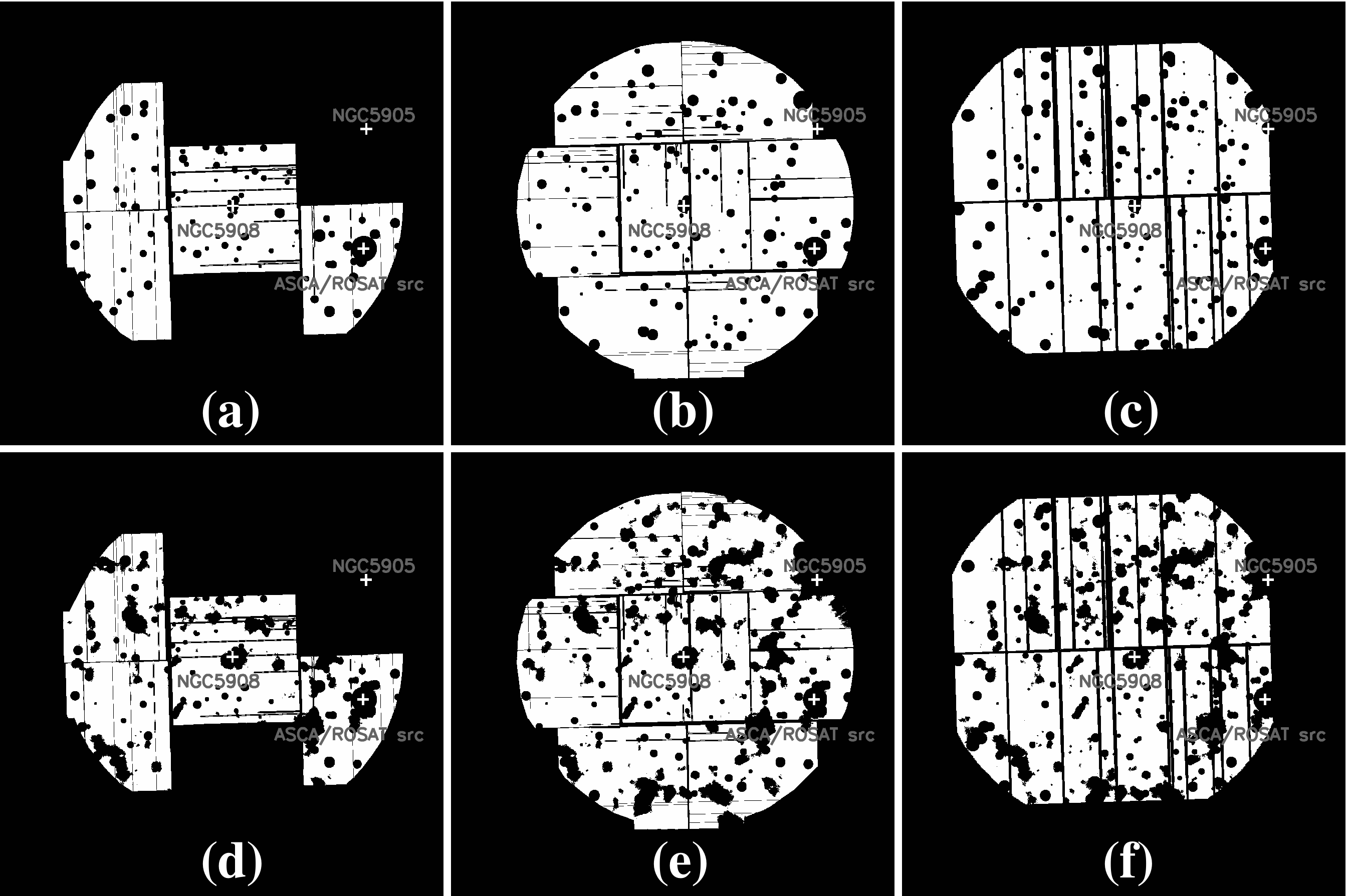,width=1.0\textwidth,angle=0, clip=}
\caption{\emph{Upper panels}: a combination of masks of detected point-like sources in the soft and hard bands for MOS-1, MOS-2, and PN. \emph{Lower panels}: a combination of point source masks in the upper panels and the masks of prominent diffuse soft X-ray (0.5-1.25 keV) features shown in Fig.~\ref{fig:images}a. The center of the most prominent features, i.e., NGC~5908, NGC~5905, and the brightest X-ray point sources detected by \emph{ASCA/ROSAT}, are marked with white pluses.}\label{fig:masks}
\end{center}
\end{figure*}

In order to analyze the faint diffuse X-ray emission, a careful detection and subtraction of point-like X-ray sources is needed. We perform source detection in two bands: 0.3-1.25~keV and 2.0-7.2~keV, using the SAS task \emph{cheese-bands}. The energy range 1.25-2.0~keV is not included in source detection, because there are some prominent instrumental lines in this band (see below; Fig.~\ref{fig:bckspec}). We adopt a point source detection limit of $L_{\rm 0.3-7.2~keV}=1.9\times10^{38}\rm~ergs~s^{-1}$ which corresponds to about 10~counts collected by MOS-1, MOS-2, and PN in 0.3-1.25~keV and 2.0-7.2~keV band. This detection limit is set to create masks in order to remove the point sources (Fig.~\ref{fig:masks}a-c).

\subsection{Background Analysis}\label{subsubsection:BackAnalysis}

For analysis of faint extended X-ray emission, it is important to decompose different background components, which have different spatial and spectral properties. We do background analysis mainly following the steps described in the ESAS background analysis threads (\url{http://xmm.esac.esa.int/sas/current/documentation/threads/esasimage_thread.shtml}).

After removing the known electron noises (such as bright pixels) and photons from out-of-time events in previous steps, the remaining \emph{XMM-Newton} EPIC background is in general comprised of two major components: the instrumental and sky backgrounds. The former has two components: the quiescent particle background (QPB) and the induced soft proton (SP) background. The latter is mostly removed in the previous step of flare filtering, but there could be some residual emission left. The sky background can be modeled with three components (e.g., \citealt{Li08,Li15b}): a low temperature ($\approx0.1\rm~keV$) unabsorbed thermal plasma representing emission from the local hot bubble, a $\approx0.3\rm~keV$ thermal plasma with Galactic foreground absorption from the Milky Way halo, and a power law with a photon index of 1.46 from distant AGN. Sometimes, there could be an additional component from solar wind charge exchange (SWCX), which is typically modeled with several separated emission lines (e.g., \citealt{Anderson16}). Including these lines in the model, however, indicates negligible contribution from them in our \emph{XMM-Newton} observation of NGC~5908. We thus do not include this component in the background modeling.

\begin{figure}
\begin{center}
\epsfig{figure=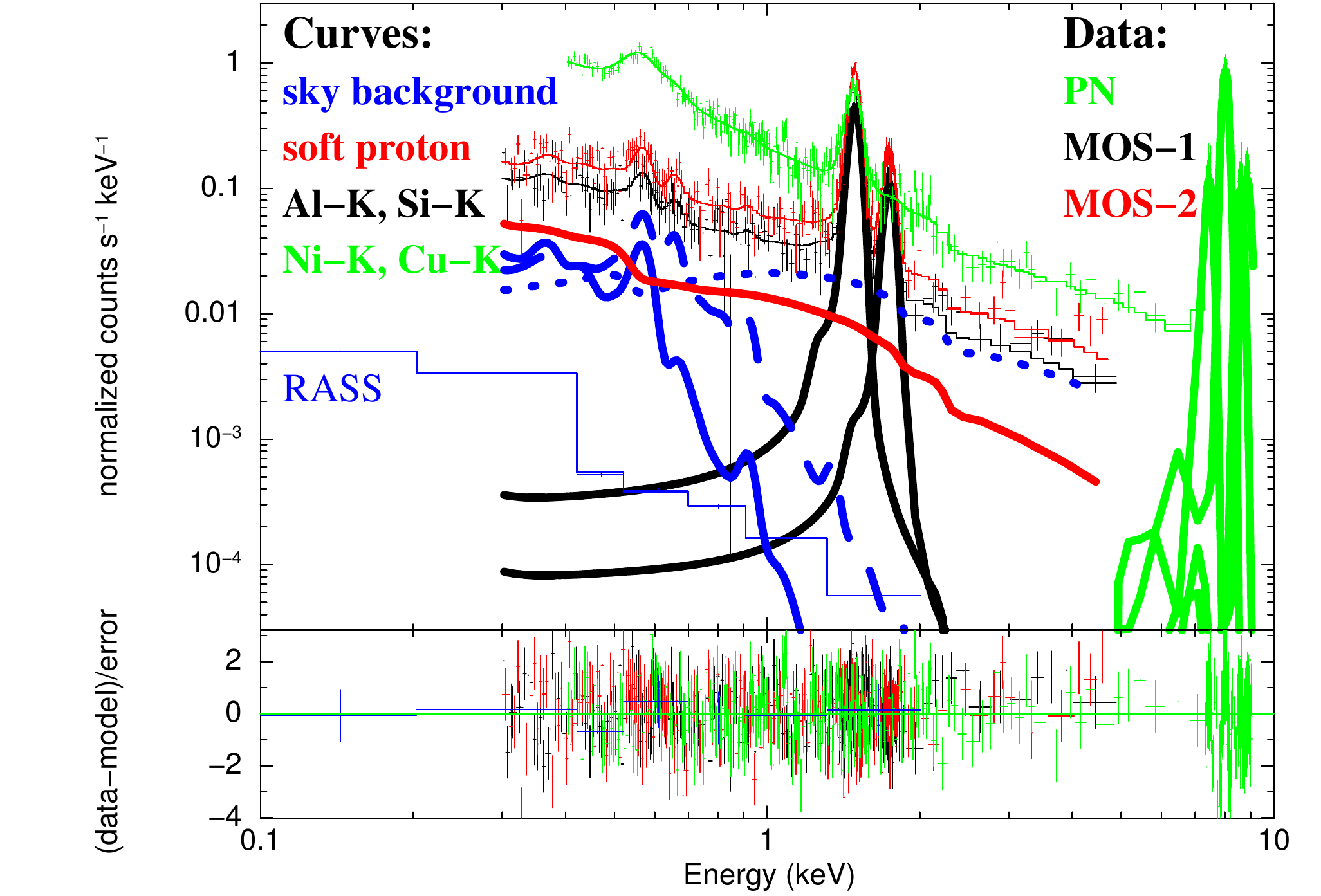,width=0.8\textwidth,angle=0, clip=}
\caption{\emph{XMM-Newton} background spectra extracted from a $r=4^\prime-12^\prime$ annulus after removing point sources and prominent diffuse X-ray features. The \emph{ROSAT} all sky survey (RASS) spectrum extracted from a $0.2^\circ-1^\circ$ annulus centered at NGC~5908 is also included in order to help constraining the sky background (blue data points and thin solid curve). Curves of different model components are scaled to the MOS-1 spectrum, except for the Ni and Cu~K$\alpha$ lines, which are scaled to the PN spectrum. Colored data points with error bars are from MOS-1 (black), MOS-2 (red), and PN (green), respectively. Colored curves denote different background model components: sky background including the local hot bubble (blue solid), the Galactic halo (blue dashed), and the distant AGN (blue dotted) components, soft proton (red), Al-K$\alpha$ and Si-K$\alpha$ instrumental lines (two black gaussian lines), and Ni-K$\alpha$ and Cu-K$\alpha$ lines of PN only (four green gaussian lines).}\label{fig:bckspec}
\end{center}
\end{figure}

Following the ESAS background analysis threads, we do background analysis in two steps. In the first step, we first extract spectra from the entire FOV using the SAS tools \emph{mos-spectra} and \emph{pn-spectra}, after removing the bright point sources using the mask created in \S\ref{subsubsection:PointSource}. The model particle background spectra and images are created with the SAS tools \emph{mos\_back} and \emph{pn\_back} in five bands: 0.3-0.5~keV, 0.5-1.25~keV, 1.25-2~keV, 2-4~keV, 4-7.2~keV. In addition, we also obtain a spectrum in the same direction from the \emph{ROSAT} All-Sky Survey (RASS) along with the appropriate spectral response matrix. This RASS spectrum help us to constrain the non-detector background model components.

We model the \emph{XMM-Newton} background spectra with the sky background models plus a thermal plasma component representing the hot gas emission from the halo of NGC~5908 and some extended background sources, as well as a broken power law representing the SP component. We also add a few Gaussian lines to account for the instrumental Al K$\alpha$ and Si K$\alpha$ lines in the MOS, and Al K$\alpha$ and Cu lines in the PN. All the cosmic components (sky background and source component), except for the local hot bubble, are subjected to the Galactic foreground absorption. The fitted background spectra are similar as shown in Fig.~\ref{fig:bckspec}, with the source component being removed (see below). 

We use the fitted SP background model to create SP background images, using the SAS tools \emph{sp\_partial} and \emph{proton}. We next combine images from different instruments using the SAS tool \emph{comb}. This tool renormalizes the exposure maps of different instruments to those of MOS-2 with a medium filter. We finally create QPB+SP-background-subtracted, exposure-corrected, and adaptively smoothed images with the SAS tool \emph{adapt} (similar as shown in Fig.~\ref{fig:images}a, see below). 

The source spectrum is comprised of different components and cannot be well constrained with the above background analysis. We therefore create a mask of prominent extended features based on the 0.5-1.25~keV flux image. The threshold of this mask is set to remove the extended features as bright as the halo of NGC~5908. We combine this extended source mask with the point source mask to create a new mask, which includes only the sky area dominated by background emission (e.g., Fig.~\ref{fig:masks}b). In the spatial and spectral analyses (\S\ref{subsection:SpatialAnalysis}, \ref{subsection:SpecAnalysis}), we will adopt this mask to the source, background, and exposure maps, in order to filter both the detected point-like sources and the prominent extended X-ray features.

In the second step, we extract and analyze the ``clean'' background spectra from a $r=4^\prime-12^\prime$ annulus, using the same methods as discribed above. All the point sources and extended X-ray features have been removed, and the spectra are fitted only with the background components (no source component). The resultant fitted background spectra and the final flux images are shown in Figs.~\ref{fig:bckspec} and \ref{fig:images}a.

\subsection{Spectra Extraction}\label{subsubsection:SpecExtraction}

We extract the EPIC source spectra using the tools detailed in \citet{Li15b}. In particular, the QPB background spectra are extracted from the Filter Wheel Closed (FWC) data, which are calibrated in the same way as the source events.
These FWC background spectra are renormalized in order to match the 10-12~keV (for MOS; 12-14~keV for PN) counts number of the source spectra. The instrumental lines are included in the FWC spectra, so the Gaussian lines in the background spectra can be removed in spectral analysis. 
For the PN source spectrum, we also subtract the expected fraction of OoT events (6.3\% for ``Full Frame'' mode).
Because the source events are already weighted with the inverse effective area (\S\ref{subsubsection:DataCalibration}), the ARFs are generated on the optical axis.
We further add a typical systematic error of 10\% to the counts rate in order to account for the calibration bias between different instruments \citep{Li15b}.

\section{Point sources and prominent extended X-ray features}\label{section:PointSrcExtendFeature}

\subsection{X-ray bright point-like sources}\label{subsection:PointSrc}

In this section, we present spectral analyses of two X-ray bright point like sources: the AGN of NGC~5908 and J151525.44+552054.96 located at the southwest corner of the FOV (Fig.~\ref{fig:pointsrcimg}).

\begin{figure*}
\begin{center}
\epsfig{figure=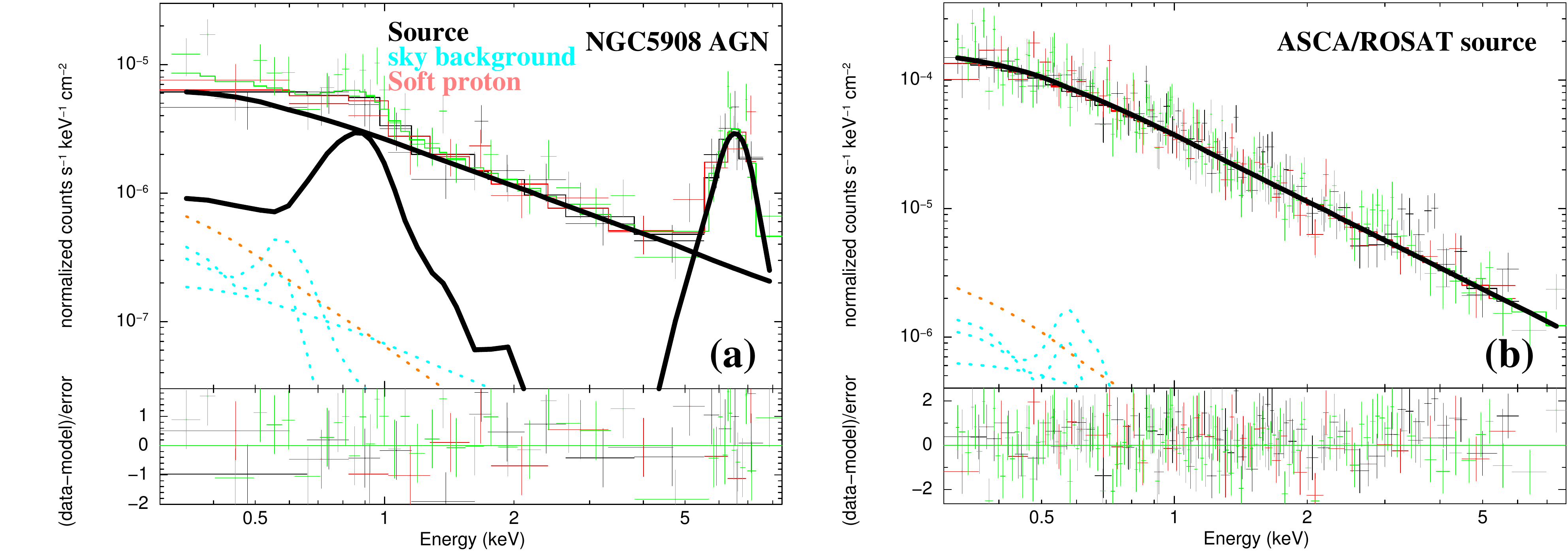,width=1.0\textwidth,angle=0, clip=}
\caption{\emph{XMM-Newton} spectra of two bright point-like sources in the FOV: (a) The nucleus of NGC~5908; (b) J151525.44+552054.96 located at the southwest edge of the FOV (Fig.~\ref{fig:pointsrcimg}c, \ref{fig:images}a). The color dotted curves represent the sky and soft proton background components, while the thick black solid curves represent the source emissions. The spectra of the AGN of NGC~5908 is well fitted with a $kT=0.8\rm~keV$ thermal plasma (APEC), a power law with photon index $\Gamma=1.26$, and a gaussian line centered at 6.53~keV ($\chi^2/d.o.f.=61.82/61$). The spectra of J151525.44+552054.96 is well fitted with a single power law with a photon index $\Gamma=1.74$ ($\chi^2/d.o.f.=235.48/237$).}\label{fig:pointsrcspec}
\end{center}
\end{figure*}

The \emph{XMM-Newton} spectra of the AGN of NGC~5908 are extracted from a circular region centered at the nucleus and with a radius of $0.2^\prime$ (Fig.~\ref{fig:pointsrcimg}b). The AGN spectra can be fitted with three components (Fig.~\ref{fig:pointsrcspec}a). The $\Gamma=1.26$ photon index of the power law is significantly smaller than those of Compton-thin AGN ($\Gamma=1.8$), indicating that the spectrum is largely due to reflected X-rays from the accretion disk (e.g., \citealt{Tozzi06}), consistent with the presence of the strong nearly neutral Fe~K$\alpha$ line (at $\approx6.53\rm~keV$) with an apparent line width of $\approx0.5\rm~keV$. In addition, a thermal component with a temperature of $\approx0.8\rm~keV$ contributes mostly to the soft X-ray emission at $\lesssim1.5\rm~keV$. This component probably represents hot gas in the galaxy's nuclear region. The total background-subtracted 0.3-8~keV luminosity of these three components is $2.1\times10^{40}\rm~ergs~s^{-1}$. The possibly extended thermal component only contributes to $\lesssim3\%$ of the total luminosity. 


The X-ray brightest point-like source J151525.44+552054.96 is located at the southwest edge of the FOV. The PSF at this location is much more extended than the on-axis one. Therefore, we adopt an elliptical region with a major/minor axis of $0.45^\prime$/$0.3^\prime$ to extract the X-ray spectra of the source (Fig.~\ref{fig:pointsrcimg}c), which are well fitted with a single power law with a photon index of $\Gamma=1.74_{-0.07}^{+0.10}$ and an absorption column density of $N_{\rm H}=0.72(<2.35)\times10^{20}\rm~cm^{-2}$ (Fig.~\ref{fig:pointsrcspec}b). The apparent 0.3-8~keV luminosity is $2.7\times10^{33}(d/{10\rm~kpc})^2\rm~ergs~s^{-1}$. The source was previously detected in X-ray with \emph{ASCA} and \emph{ROSAT}, but was not yet identified. We find a potential point-like optical counterpart that is located only $\approx0.7^{\prime\prime}$ away from the X-ray centroid. The apparent SDSS magnitudes of this optical counterpart are: $u=21.726\pm0.178$, $g=20.722\pm0.030$, $r=20.522\pm0.036$, $i=20.403\pm0.046$, $z=19.981\pm0.118$. Therefore, it is relatively bright in UV and IR compared to the Sun. If the u-band absolute magnitude equals to the Sun's, the source should then be located at $\approx10\rm~kpc$ from us. In this case, the source could be a foreground stellar source in the Milky Way, consistent with the low absorption column density as indicated by the soft X-ray spectra.  

\subsection{Extended soft X-ray features}\label{subsection:ExtendedFeature}

In addition to the extended galactic corona around NGC~5908 (Fig.~\ref{fig:images}b), we also find some prominent extended soft X-ray features in the FOV of the \emph{XMM-Newton} observation (Fig.~\ref{fig:images}a).


The companion galaxy NGC~5905 is located at the edge of the FOV. The diffuse soft X-ray emission from this galaxy is also detected, but overlaps with a background group of galaxy SDSSCGB~42645. We do not find evidence for any highly significant extended X-ray features which may arise from the IGM of the NGC~5905/5908 system. Because the diffuse X-ray emission is directly detected only to $r\approx2^\prime$ from the center of NGC~5908 (Fig.~\ref{fig:profile}; compared to the $\approx13^\prime$ separation of NGC~5905/5908), we could regard NGC~5908 as isolated in the study of its extended hot halo. This is consistent with the non-detection of the intragroup medium in the surrounding area \citep{Pisano04}.

There are two other galaxies in the FOV which have some associated diffuse X-ray emission features: J151545.14+552624.3 ($z=0.35912$) and J151538.17+552919.0 ($z=0.70247$). Each is the Brightest Cluster Galaxy (BCG), so their diffuse X-ray emission may alternatively be from the ICM.


Except for NGC~5905/5908, the most prominent extended X-ray features in the FOV are associated with two galaxy clusters at similar redshifts ($z=0.4553$ for WHL~J151600.8+551908 and $z=0.4551$ for WHL~J151549.1+552613). The centers of these two clusters are separated by $\approx3.5^\prime$, or $1.6\rm~Mpc$ in projection. There is a faint diffuse X-ray filament connecting the extended X-ray emission of the two clusters, indicating that they are interacting with each other.

There is another X-ray bright galaxy cluster J151741.5+552730 ($z=0.5170$) located to the east of NGC~5908.


In addition to the core of NGC~5905 and NGC~5908, we have also identified three X-ray bright AGNs; two of them, J151545.1+553518 ($z=1.65156$) and J151800.17+551331.0 ($z=2.37531$), are at the edge of the FOV, while the other one, J151552.0+552201 ($z=2.21039$), lies between the double cluster.


All the above extended soft X-ray features lie well beyond $2^\prime$ from the nucleus of NGC~5908, and thus should not affect our results on the hot halo of this galaxy.

\end{document}